# Use of the European Data Grid software in the framework of the BaBar distributed computing model

D. Boutigny
*Laboratoire de Physique des Particules,  F-74941, Annecy-le-Vieux, France*
D.H. Smith
*University of Birmingham, Birmingham B15 2TT, United Kingdom*
E. Antonioli, C. Bozzi, E. Luppi, P. Veronesi
*Universit‡ di Ferrara, Dipartimento di Fisica and INFN, I-44100 Ferrara, Italy*
G. Grosdidier
*Laboratoire de l'AccÈlÈrateur LinÈaire, F91898 Orsay, France*
D. Colling, J. Martyniak, R. Walker
*University of London, Imperial College, London SW7 2BW, United Kingdom*
R. Barlow, A. Forti, A. McNab
*University of Manchester, Manchester M13 9PL, United Kingdom*
P. Elmer
*Princeton University, Princeton, New Jersey 08544*
T. Adye
*Rutherford Appleton Laboratory, Chilton, Didcot, Oxon OX11 0QX, United Kingdom*
B. Bense, R. D. Cowles, A. Hasan, D. A. Smith
*Stanford Linear Accelerator Center, Stanford, California 94309*

*On behalf of the BaBar Computing Group*

We present an evaluation of the European Data Grid software in the framework of the BaBar experiment. Two kinds of applications have been considered: first, a typical data analysis on real data producing physics n-tuples, and second, a distributed Monte-Carlo production on a computational grid. Both applications will be crucial in a near future in order to make an optimal use of the distributed computing resources available throughout the collaboration.

## 1. THE BABAR COMPUTING FRAMEWORK

The BaBar experiment [1], located at the Stanford Linear Accelerator Center (SLAC) has been taking physics data since 1999. Its original computing model did not take into account any Grid concept.

Since 2001, BaBar has been evolving toward a distributed computing model based on a multi-Tier approach similar to what is foreseen for the LHC experiments.

- Tier-A sites are major computing centers holding all or a significant fraction of the experimentsí data. They provide computing access to any collaborator. They are the primary source for other sites to copy data from.
- Tier-B are intermediate centers holding the data corresponding to dedicated physics channels, they were foreseen in the model but were not really deployed.
- Tier-C are individual institutes or university with limited computing resources. They have a copy of some physics data in direct relation to the local analyses.

At the moment BaBar has Tier-A at SLAC, CCIN2P3, RAL, INFN-Padova and FZK/GridKA. The INFN-Padova center is specialized in data reprocessing while others are dedicated to analysis.

The Monte-Carlo production is distributed between 23 sites in the collaboration. Each site has a local contact person in charge of producing data and transfer to SLAC. More than 1.6 billion Monte-Carlo events have been produced up to now.

Presently BaBar supports two different event formats; one is based on the Object Oriented database system Objectivity [2] and the other is using ROOT I/O [3]. In the future the Objectivity format is to disappear from the event store but will remain for the database holding constants and data taking conditions.

In such a distributed environment, the Grid concept is very appealing in order to make the best possible use of our resources and in order to provide a simple framework for data analysis where physicists do not need to worry where the data is located.

Nevertheless, we need to keep in mind that BaBar is a running experiment producing first class physics results, therefore the introduction of a Grid technology should not break the existing software and should not disrupt the analysis effort in any way.

## 2. THE BABAR GRID INFRA-STRUCTURE

Several BaBar institutes are also involved in LHC experiments and were already participating in the European Data Grid (EDG) project [4]. Some were involved in the EDG test-bed deployment. It was therefore quite natural for BaBar to evaluate the EDG middleware as a basis for BaBar Grid implementation [5]. Since 2002 BaBar is officially part of the EDG Work Package 8 (WP8) devoted to HEP applications.

A global view of the BaBar Grid configuration as of March 2003, is illustrated in Figure 1.

### 2.1. The BaBar Virtual Organization

The basis of the BaBar Grid is a Virtual Organization (VO) maintained in Manchester. This VO accepts certificates issued by any EDG certification authority as well as by the US Department Of Energy (DOE). Once issued, users copy the certificate header into a special identification file in the SLAC AFS system. A cron job checks on a regular basis the presence of these files and updates the VO accordingly. The presence of a given set of ACLs on their identification file is the proof that the user is registered in BaBar.

### 2.2. The Resource Broker

BaBar Grid is sharing the EDG and GridPP [9] test-bed Resource Broker (RB), maintained at Imperial College. This RB has been regularly upgraded to follow successive EDG releases.

Even in its latest version (EDG V1.4.8) the RB software was known to be unstable. The problem was that the brokering mechanism relies on a dynamic information system or Meta Directory Service (MDS) that is unable to handle disappearing or reappearing unstable sites. Two possible solutions were implemented to overcome this problem:

1. A set of monitoring tools scrutinized the registered sites and would stop and restart the MDS each time one of them changed between the present and absent state. This solution was unfortunately unable to handle cases where sites are rapidly fluctuating between the two states. In these cases the only solution was to remove the site from the list permanently, which required manual intervention.
2. The dynamic MDS published information into a static one called the Berkeley Database Information Index (or BDII). The BDII updates its information every 10 minutes from the dynamic MDS, however if information about a site is not available it will continue to publish the most recent information that it has. This second solution is the one recommended by EDG and should lead to a better stability.

Both solutions were tested and results will be presented later in this paper.

The CNAF (Bologna) Resource Broker has also been used during the Grid Monte-Carlo production tests.

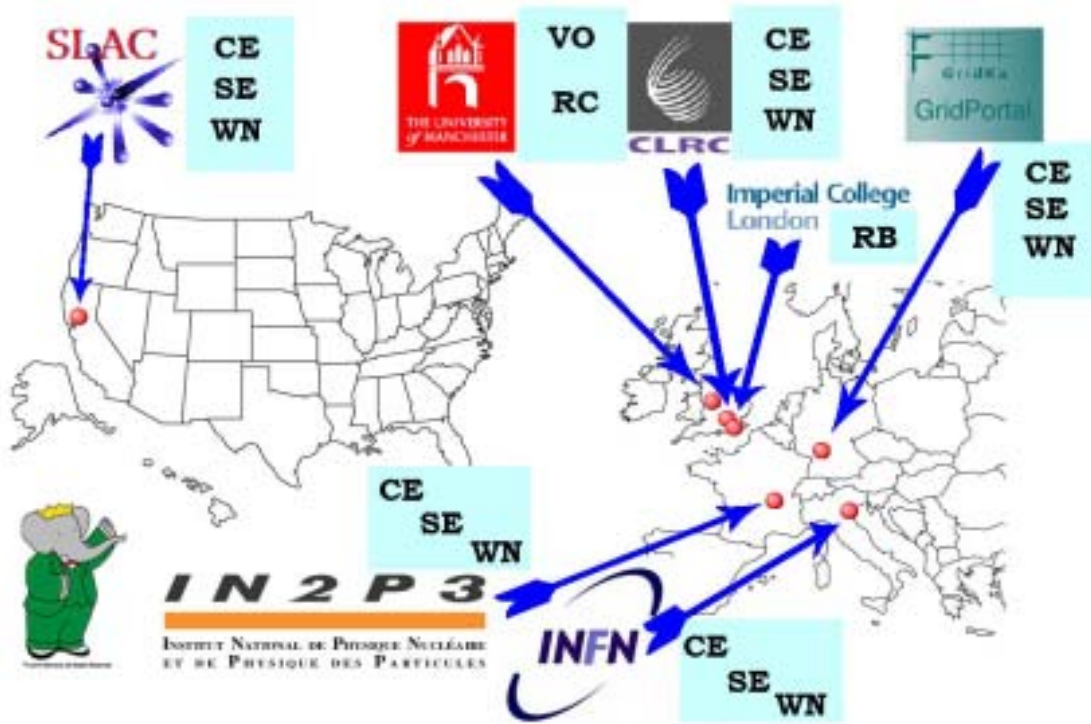

Figure 1: The BaBar Grid configuration as of March 2003

**Insert PSN Here**

## 2.3. The Computing Elements, Storage Elements and Worker Nodes

Each EDG test-bed sites maintains one or more Computing Element (CE) connected to a farm of Worker Nodes (WN) and one or more Storage Elements (SE).

The BaBar Grid has been re-using part of this existing infrastructure in the European sites.

We have added a Replica Catalog (RC), maintained in Manchester in order to keep track of the files registered in the SE. In principle the RC is coupled to a Grid Data Mirroring Package (GDMP) system in charge of replicating the files between different SE. BaBar has deliberately chosen to not install any GDMP system and to manage the replication by hand. We prefer, and look forward to, using the new Replica Location Service (RLS), which is announced for the version 2.0 of the EDG software. The Replica Catalog and Replica Manager (RM) software have several known limitations in term of scalability but we judged their functionalities crucial for the present evaluation.

## 2.4. The SLAC setup

SLAC is not part of the EDG test-bed infrastructure, the EDG CE, SE and WN as well as the User Interface (UI) had to be installed from scratch. The SLAC security rules impose that all the farm nodes used as WN are located behind a firewall in order to be isolated from the Internet, and that there is neither inbound nor outbound connectivity. The EDG setup had to be modified to fulfill this rule. The scripts in charge of the communication between the WN and the RB have been split in order to direct any such communication through the CE which is accessible from the Internet.

The fact that the official EDG release was only available under the RedHat 6.2 system was a very serious concern from the security viewpoint. Two other SLAC specificities were of concern:

- The SLAC LRMS (Batch Scheduler) is LSF, and was poorly tested inside of the Globus framework, the job monitoring was shaky, probably due to some additional NFS instabilities.
- The home directories are managed through AFS, and security requirements led us to use real user accounts, not pool users as for other sites, to allow for easier job/user traceability. This was a major culprit because there was no way to easily acquire an AFS token, which was required to give full access to the ".gass_cache" directory to the Globus job manager. Two different workarounds were used in turn to allow for this access: we tried to move the .gass_cache area into a NFS file system (together with a soft link for each user), but were hit then by the NFS instability issues already mentioned. We then decided to use the GSSKLOG software provided by the Globus team on an experimental basis to acquire an AFS token through GSI authentication. The latter solution was safer, although it required a modification of the interface between the gatekeeper process and the job manager instance to communicate with the AFS server to get the token.

## 3. THE ANALYSIS USE CASE

A standard BaBar analysis job reads input data from a set of event collections and produces either new collections containing a subset of the initial ones, or a set of n-tuples.

The executable is built locally from a standard BaBar release; its input is a rather complicated set of inter-dependant Tcl files spread in the release structure. A feature of the BaBar Framework mechanism allows users to produce a complete configuration dump in the form of a single Tcl file.

A BaBar executable has some external dependencies on shared libraries (like Objectivity, Root etcÖ) and on configuration data files which are kept in the release structure. In principle, it is possible to package all these dependencies and to distribute them on the Grid, but for simplicity we assumed in a first step, that a standard release structure was available in the Grid target sites.

### 3.1. Test description

The test consisted of the following steps:
1. An executable and an input Tcl file is prepared
2. The executable is distributed to the SE close to a set of CE and registered to the RC under a logical file name.
3. A Job Description Language (JDL) script is prepared in such a way to:
    a. Send the Tcl file and a wrap-up script to the CE through the Sandbox mechanism.
    b. Select a CE which holds a close SE where the executable is available
    c. Run the job on the selected CE.
    d. Return the output to the submission site through the Sandbox mechanism.

The Sandbox mechanism allows to easily and safely transmit input and output files between the CE and the submission site.

The wrap-up script is in charge of:
1. Setting-up the directory structure on the WN in order to be compatible with the BaBar setup.
2. Copy the executable from the close SE to the WN
3. Run the executable
4. Transfer the output n-tuple to the close SE and register it into the RC

With EDG version 1.4 it is possible to make all these JDL and wrap-up script fully generic. They are able to run on any node having a standard BaBar release structure, and to discover resources at run time, such as the SE address, the SE mount point and the executable physical file name.

## 3.2. Test results

In order to determine the success rate we sent several bunches of 200 jobs to the Grid, with an executable typical of a real analysis application. We define the success rate as the fraction of jobs successfully completed with the full output returned to the submission site and a valid n-tuple stored and registered on the SE.

With the dynamic MDS the success rate varied from 55% to 75%. 98% of the failing jobs were due to the RB unable to match the requested resources to any CE, confirming the known limitations of the dynamic MDS.

With the static MDS (BDII) the success rate went up to 99%, the failing 1% corresponded to jobs lost by the RB. This last result has been obtained on a total of 800 jobs submitted.

There is another scaling issue in the Condor-G component setting a maximum of 512 jobs present at the same time on the RB. One full bunch of 200 jobs has been impacted by this limit and has been lost. This lost is not accounted for in the results presented above.

The RB system is sometimes in a faulty state due to a corruption in the PostGres database system. When this occurs, the database should be cleared up and all the jobs present in the RB are lost.
It is expected that these limitations will be removed in future EDG releases.

## 3.3. Conclusions on the analysis application

The results obtained during the test were very encouraging. It was possible to demonstrate that the EDG system is usable in an analysis environment. The success rate is compatible with a production system.

The fact that it is possible to write generic submission scripts and to discover the resource characteristics at run time, is very promising. This allows us to envisage how we could implement rather sophisticated tasks on the Grid.

On the other hand we clearly see obvious scalability issues, like the 512 job limit or on the size of the RC. We expect that further releases or other Grid implementations will solve these problems.

## 4. MONTE-CARLO PRODUCTION

In a second application, the EDG software has been used to test the production of Monte-Carlo events for the BaBar experiment in a Grid environment.

A farm of the INFN-Grid test bed (in Ferrara), with EDG software installed, has been specially customized to install an Objectivity database and run BaBar simulation software. Several test runs of thousands of simulated events have been successfully submitted by using the standard EDG tools interfaced with the already existing Babar simulation production tools. The INFN-Grid Resource Broker at CNAF (Bologna) was used.

The production log files were retrieved through the Sandbox mechanism. The output of the simulation, consisting of larger files (15MB per run) in ROOT I/O format, were copied to the closest Storage Element and retrieved with simple EDG commands without passing through the Resource Broker.

## 4.1. Software Configuration

The BaBar Monte-Carlo simulation software consists of an executable *Moose,* which reads conditions and background triggers from an Object-Oriented database (Objectivity), generates events, propagates particles through the detector, adds background from real data by using random triggers and finally performs the full event reconstruction. Moose is driven by a command file written in Tcl, which sets the sequence of modules to be executed, and which allows users to specify the generation parameters, and to choose detector conditions and calibration constants. The output of the simulation is normally stored in Object-Oriented databases. Both the Objectivity Database and Moose are not part of the EDG software releases.

The standard configuration for an EDG Testbed farm has been modified in Ferrara by adding a pool of BaBar accounts (babar001, babar002,...) and creating some BaBar-specific LCFG (Local ConFiGuration Server) configuration files, similar to the ones used by other experiments (i.e. CMS, Alice, Atlas, ...). We enabled the access to the farm for users registered with the BaBar VO. In this way, each member of BaBar VO is able to submit jobs to the Ferrara farm through the Italian Resource Broker located at CNAF.

The BaBar simulation software was installed on a separate data server (running the Solaris operating system), which sits outside the Grid environment, and which exports via NFS to the Grid Elements of the farm.

A dedicated BaBar software release was needed and built to run Moose on clients based on the Linux RedHat 6.2 operating system, as required by the current EDG software, and to write files in ROOT I/O format instead of Object Oriented databases. Detector conditions and calibration constants were nevertheless needed in the form of an Object Oriented database, which was therefore installed on the Solaris data server.

## 4.2. Job submission and output retrieval

The submission of BaBar Monte-Carlo runs on the Grid was accomplished by using standard JDL commands, wrapped in very simple scripts containing the standard EDG commands to submit jobs, to check their status and retrieve the results of the simulation. These scripts also allow users to specify a run range, the detector conditions and the type of events to be generated.

For each run a job is created and sent to the Ferrara Computing Element (grid0.fe.infn.it) through the CNAF

Resource Broker. The output file is then transferred to the closest Storage Element (grid2.fe.infn.it). We successfully submitted jobs, each of them producing 2000 Monte Carlo events, as in the standard simulation production activity.

ROOT I/O files were successfully copied and registered to the Storage Element through the BaBar Replica Catalog located in Manchester.

The more complex software tools which manage Monte Carlo simulation jobs in BaBar have also been partially tested and validated, and will be used in future.

### 4.3. Conclusions on Monte-Carlo production

The BaBar simulation software is now being checked for package dependencies in order to build an RPM that can be integrated in a future EDG software release, which can in turn be distributed and automatically installed in other EDG sites. Therefore, we will be able to export this model to all BaBar and also non-BaBar sites, provided that an Objectivity DataBase can be accessed efficiently to read conditions and calibration constants.

The global management of the production of simulated events in many remote sites already exists in BaBar [6], and can be optimized by porting it to a Grid environment. In this respect, we are evaluating already existing tools (e.g. Genius [7]) in order to embed all user services in a single interface.

### 5. SUMMARY AND OUTLOOK

The EDG software has been evaluated by the BaBar experiment for analysis and Monte-Carlo production applications. In both cases we have been able to setup a realistic environment demonstrating that both applications can be run on Grid. We conclude that all the necessary functionalities are present but some scalability issues remain.

In the near future we intend to test the Virtual Data Toolkit (VDT) [8] in the BaBar environment and to explore the inter-operability between European and US Grid implementations.

### Acknowledgments

The authors wish to thank Nadia Lajili and Fabio Hernandez from CCIN2P3 for their help in setting up the BaBar Grid infrastructure.